\documentclass[pra,twocolumn,showpacs,preprintnumbers,amsmath,amssymb]{revtex4}
\usepackage{mathrsfs}
\usepackage{txfonts}
\usepackage{graphicx}% Include figure files
\usepackage{dcolumn}% Align table columns on decimal point
\usepackage{bm}% bold math
\usepackage{amsmath}
\begin{document}

%\preprint{APS/123-QED}

\title{Quantum information transfer in a coupled resonator waveguide }% Force line breaks with \\

\author{Peng-Bo Li$^1$}
\author{Ying Gu$^1$}
%\email{}
\author{ Qi-Huang Gong$^{1}$}
\author{ Guang-Can Guo$^{1,2}$}
\affiliation{$^{1}$State Key Laboratory for Mesoscopic Physics,
Department of Physics, Peking University, Beijing 100871, China\\
$^{2}$Key Laboratory of Quantum Information, University of Science
and Technology of China, Hefei 230026, China}

%\date{April 9, 2008}

\begin{abstract}
We propose an efficient scheme for the implementation of quantum
information transfer in a one-dimensional coupled resonator
waveguide. We show that, based on the effective long-range
dipole-dipole interactions between the atoms mediated by the cavity
modes, Raman transitions between the atoms trapped in different
nodes can take place. Quantum information could be transferred
directly between the opposite ends of the coupled waveguide without
involving the intermediate nodes, via either Raman transitions or
the stimulated Raman adiabatic passages. Since this scheme, in
principle, is a one-step protocol, it may provide useful
applications in quantum communications.
\end{abstract}

 \maketitle

\section{INTRODUCTION}
The transfer of quantum states from one place
to another is an important goal in the field of quantum information
science for distributing and processing information
\cite{quantum_information,nature-453-1023}. To accomplish this task,
several approaches have been employed. For long distance quantum
communications, optical systems such as cavity QED system
\cite{Kimble,Sci298} are used to transfer states from one node to
another through photons transmitting in a fiber
\cite{prl-78-3221,prl-98-193601}. For the case of short distance
quantum communications, spin chains are proposed
\cite{prl-91-207901}. In spin chains, single spin addressing is
difficult because the spatial separation between neighboring spins
is very small. Thus the control over the couplings between the spins
or over individual spins is very hard to achieve. Therefore, these
protocols based on spin chains have some drawbacks, which impair the
performance for quantum information transfer (QIT). Recently
coupled-resonator waveguide has attracted great interests
\cite{Nature-physics-2-849,
Nature-physics-2-856,Prl-99-160501,Prl-99-103601,Prl-99-183602,Prl-101-100501,Prl-101-246809}.
We have proposed a protocol for generating atomic cluster states
using coupled resonators for one-way quantum computation \cite{li}.
Coupled resonator waveguide has the advantage of easily addressing
individual lattice sites with optical lasers. Furthermore, the atoms
trapped in the resonators can have relatively long-lived atomic
levels for encoding quantum information. Therefore, it is desirable
to develop a technique for implementing short distance quantum
communications in a coupled-resonator waveguide.

In this work, we propose a scheme for the implementation of QIT
between three-state atoms trapped in a one-dimensional
coupled-resonator waveguide. We first demonstrate the coupled system
can be reduced to an effective $\Lambda$ configuration which
supports Raman transitions between the first atom and the end one.
Then we utilize this protocol to implement short distance quantum
communications. This proposal exploits the effective long-range
dipole-dipole interactions mediated by the cavity modes between the
atoms. The nonlocal interactions combined with lasers are utilized
to induce Raman transitions between the atoms trapped in the two
ends of the waveguide via the exchange of virtual cavity photons.
Quantum states can be transferred directly from the first node to
the end one within the one-dimentional coupled-resonator waveguide,
through either Raman transitions or the stimulated Raman adiabatic
passages (STIRAP)\cite{RMP-70-1003}. Up to our knowledge, this is
the first proposal for QIT in a network using coupled resonators,
which should provide very interesting applications in the field of
quantum information processing, such as entanglement distribution,
teleportation \cite{Prl-70-1895}, and distributed quantum
computation \cite{pra-56-1201}. Experimentally this protocol could
be realized with the state-of-the-art technology.

\section{QUANTUM-INFORMATION TRANSFER
IN A ONE-DIMENSIONAL COUPLED RESONATOR WAVEGUIDE}
\begin{figure}[h]
\centering
\includegraphics[bb=102 544 480 750,totalheight=1.9in,clip]{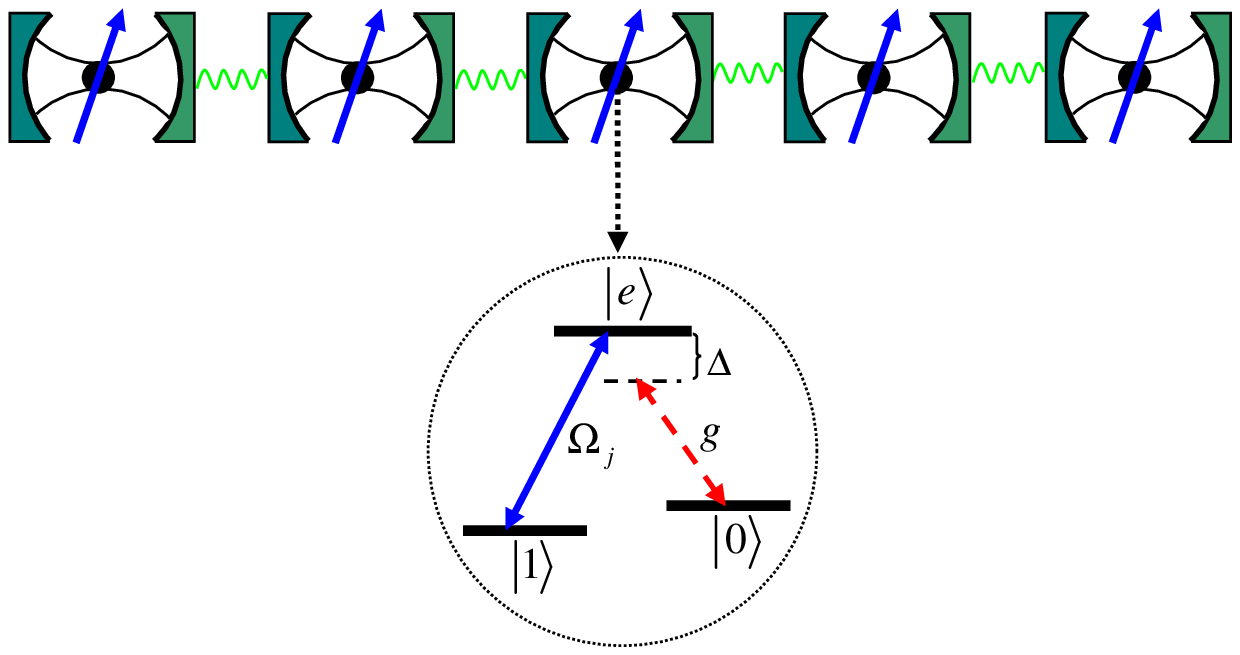}
\caption{\label{F1}Schematic diagram of a one-dimensional
coupled-resonator waveguide consisting of $N$ nodes and three-state
atoms trapped in each resonator. The transition $\vert
0\rangle\leftrightarrow\vert e\rangle$ is strongly detuned from the
cavity modes, which induces a long range interaction between the
atoms.}
\end{figure}

Consider a
one-dimensional coupled-resonator waveguide consisting of $N$ nodes,
as sketched in Fig. \ref{F1}. The coupled-resonator waveguide can be
realized in a wide range of physical systems, such as nanocavities
in photonic crystals \cite{nature-445-896}, and superconducting
transmission line resonators \cite{nature-451-664}. To implement
QIT, each node consists of a cavity and a trapped three-state atom.
Each atom has the level structure of a three-state system with two
lower states $\vert 0\rangle_j$ and $\vert1\rangle_j$
($j=1,2,...,N$) for storage of one qubit of quantum information, and
an upper state $\vert e\rangle_j$. The cavity mode is far detuned
from the atomic transition $\vert 0\rangle_j\leftrightarrow\vert
e\rangle_j$ (transition frequency $\omega_0$) with the coupling
constant $g$ and detuning $\Delta$. The transition $\vert
1\rangle_j\leftrightarrow\vert e\rangle_j$ (transition frequency
$\omega_1$) in each atom is driven resonantly with lasers (frequency
$\omega_{Lj}=\omega_1$) with Rabi frequencies $\Omega_j$.

We consider each trapped atom interacting with the cavity fields and
lasers. The Hamiltonian that describes the photons in the cavity
modes is \cite{Nature-physics-2-849,
Nature-physics-2-856,Prl-99-160501,Prl-99-103601,Prl-99-183602,Prl-101-100501}
\begin{eqnarray}
\hat{H}_c&=&\omega_c\sum^N_{j=1}a^\dag_ja_j
+J_c\sum^N_{j=1}(a^\dag_ja_{j+1}+a_ja^\dag_{j+1}),
\end{eqnarray}
$a_j$ is annihilation operator for the photon in cavity $j$, and
$J_{c}$ is the hopping rate of photons between neighboring cavities.
For convenience we introduce the notation $\mathbf{J}=(uj,0,0)$ to
denote the position of the $j$th site where $u$ is the length of the
one-dimensional crystal cell. If the periodic boundary conditions
are considered, $\hat{H}_{c}$ can be diagonalized through the
Fourier transformation. Then we obtain $\hat{H}_{c}=\sum_{k}\omega
_{k}a_{k}^{\dag }a_{k}$, where $\omega _{k}=\omega _{c}+2J_{c}\cos
k$, and $k=(2\pi m)/(Nu)$ for $m=0,1,...,N-1$. Under the rotating
wave and dipole approximations, the interaction between the atoms
and cavity fields is
\begin{eqnarray}
\hat{H}_{ac}&=&\sum^N_{j=1}g(a^\dag_j\vert0\rangle_j\langle
e\vert+a_j\vert e\rangle_j\langle 0\vert),
\end{eqnarray}
and the interaction between the atoms and lasers reads
\begin{eqnarray}
\hat{H}_L&=&\sum^N_{j=1}(\Omega_je^{-i\omega_{1}t}\vert
e\rangle_j\langle 1\vert+\mbox{H.c.}).
\end{eqnarray}
Here we add a laser to each resonator for generality, but in the
following when we discuss how to implement QIT, we in fact only
require the lasers added to the first and the last cavities be
switched on. In the interaction picture the Hamiltonian that governs
the coupled system is
\begin{eqnarray}
\label{HI} \hat{H}_I&=&\sum^N_{j=1}[\Omega_j\vert e\rangle_j\langle
1\vert+\sum_kg/\sqrt{N}\vert0\rangle_j\langle e\vert a_k^\dag
e^{i\textbf{k}\cdot\textbf{J}+i\delta_kt}+\mbox{H.c.}],\nonumber\\
\end{eqnarray}
with $\delta_k=\omega_k-\omega_0$. To further reduce the model, we
assume $\delta _{k}\gg g$ (for all $k$), then we can adiabatically
eliminate the photons from the above description \cite{Guo,method}.
By considering the terms up to second order and dropping the fast
oscillating terms, we obtain the following effective Hamiltonian
\begin{widetext}
\begin{eqnarray}
\label{H1} \hat{H}_{\mbox{eff}}&=&\sum_{j=1}^N[J_0\vert
e\rangle_j\langle e\vert+(\Omega_j\vert e\rangle_j\langle 1\vert
+\sum_{l=1}^{N}J_l\vert 0\rangle_j\langle e\vert\otimes\vert
e\rangle_{j+l}\langle 0\vert+\mbox{H.c.})],
\end{eqnarray}
\end{widetext}
with
$J_0=\sum_k[g^2/(N\delta_k)],J_l=\sum_k[g^2e^{ikl}/(N\delta_k)]$,
and the conventions $\vert e\rangle_{N+i}\langle
0\vert\equiv0,(i=1,2,...,N)$. The first term corresponds to the
level shift for each atom, the second term describes the
interactions between atoms and lasers, and the last term represents
the effective dipole coupling of trapped atoms induced by cavity
modes.
\begin{figure}[h]
\centering
\includegraphics[bb=62 402 490 758,totalheight=2.5in,clip]{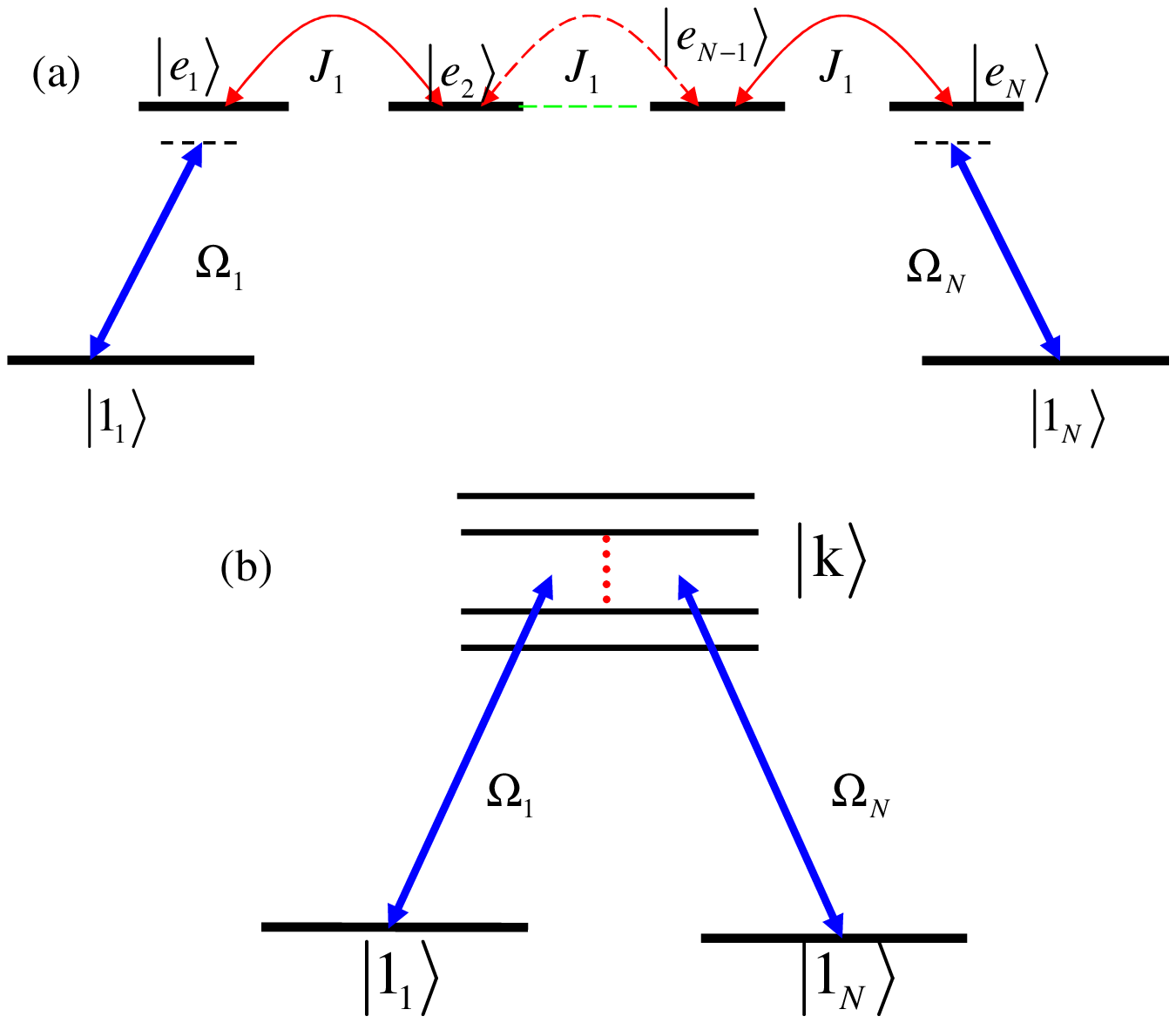}
\caption{(a) Schematic diagram of the system under the interaction
Hamiltonian (\ref{H1}). (b) Coupling configuration corresponding to
(a) in a new basis $\{\vert\textbf{1}_1\rangle,\vert
\textbf{k}\rangle...,\vert\textbf{1}_N\rangle\}$.}
\end{figure}

We introduce the states
$\vert\textbf{1}_i\rangle=\vert000...1_i...000\rangle$ and
$\vert\textbf{e}_i\rangle=\vert000...e_i...000\rangle$, which denote
that the atom at the $i$th site has been flipped to the state
$\vert1\rangle$ and $\vert e\rangle$ while other atoms stay in
$\vert0\rangle$. Assuming that only the lasers $\Omega_1$ and
$\Omega_N$ are switched on and the system initially stays in
$\vert\textbf{1}_1\rangle$, then the coupling scheme can be
schematically illustrated in Fig. 2(a). To gain more insight into
this coupling configuration, we turn to a new basis
$\{\vert\textbf{1}_1\rangle,\vert
\textbf{k}\rangle...,\vert\textbf{1}_N\rangle\}$, where $\vert
\textbf{k}\rangle=1/\sqrt{N}\sum^N_{j=1}e^{ikj}\vert\textbf{e}_j\rangle$.
Then we could diagonalize the effective dipole coupling
$V_d=\sum_{j=1}^N[J_0\vert e\rangle_j\langle
e\vert+\sum_{l=1}^{N}(J_l\vert 0\rangle_j\langle e\vert\otimes\vert
e\rangle_{j+l}\langle 0\vert+\mbox{H.c.})]$ in this subspace. The
eigenstates of $V_d$ are $\{\vert \textbf{k}\rangle,k=(2\pi
m)/N(m=0,1,...,N-1)\}$. The eigenvalues are given by
$E_k=J_0+\sum_{l=1}^N2J_l\cos(kl)$. Thus we can write $V_d$ as
$V_d=\sum_kE_k\vert \textbf{k}\rangle\langle\textbf{k}\vert$. In
such a case, the effective Hamiltonian $\hat{H}_{\mbox{eff}}$ can be
rewritten in the subspace $\{\vert\textbf{1}_1\rangle,\vert
\textbf{k}\rangle...,\vert\textbf{1}_N\rangle\}$ as
\begin{eqnarray}
\label{H2}
\hat{H}_{\mbox{eff}}&=&\sum_k[E_k\vert\textbf{k}\rangle\langle\textbf{k}\vert
+(\Omega_{1k}\vert\textbf{k}\rangle\langle\textbf{1}_1\vert\nonumber\\
&&+\Omega_{2k}\vert\textbf{k}\rangle\langle\textbf{1}_N\vert+\mbox{H.c.})],
\end{eqnarray}
with
$\Omega_{1k}=\langle\textbf{k}\vert\hat{H}_{\mbox{eff}}\vert\textbf{1}_1\rangle=\Omega_1e^{ik}/\sqrt{N}$
and
$\Omega_{2k}=\langle\textbf{k}\vert\hat{H}_{\mbox{eff}}\vert\textbf{1}_N\rangle=\Omega_Ne^{iNk}/\sqrt{N}$.
 The
schematic diagram of this coupling configuration in this new basis
is shown in Fig. 2(b), from which we see that the Hamiltonian
(\ref{H2}) describes an effective $\Lambda$ system, with two lower
states $\vert\textbf{1}_1\rangle,\vert\textbf{1}_N\rangle$ and
several upper states $\vert\textbf{k}\rangle$. Under the conditions
$E_k\gg\{\Omega_{1k},\Omega_{2k}\}$, Raman transitions can take
place between the states $\vert\textbf{1}_1\rangle$ and
$\vert\textbf{1}_N\rangle$. Through adiabatic elimination of the
states $\vert\textbf{k}\rangle$, the effective Hamiltonian
describing this case is
\begin{eqnarray}
\label{H3}
\hat{H}_{\mbox{eff}}&=&\Theta_r\vert\textbf{1}_N\rangle\langle\textbf{1}_1\vert+\mbox{H.c.},
\end{eqnarray}
with $\Theta_r=\sum_k\Omega_{1k}\Omega^*_{2k}/E_k$ the effective
Raman transition rate. This Hamiltonian describes direct Raman
transitions between the first node and the last one, assisted by the
intermediate nodes through virtual photon exchanges.

We now discuss how to implement QIT in this one-dimensional coupled
resonator waveguide. We assume that the state sender Alice is
located at the first node, and the state receiver Bob is located at
the end of the waveguide. Alice wants to transfer an unknown state
to Bob through this waveguide. To start the protocol, Alice places
the atom at the first site in the arbitrary unknown state
$\alpha\vert 0\rangle_1+\beta\vert 1\rangle_1$, while the atoms in
other nodes are prepared in the state $\vert0\rangle$. We can
describe the state of the whole system at this instant as
$\vert\Psi(0)\rangle=\alpha\vert\textbf{0}\rangle+\beta\vert\textbf{1}_1\rangle$,
with $\vert\textbf{0}\rangle=\vert000...0\rangle$. Then under the
interaction of Eq. (\ref{H3}), the state vector at the time $t$ is
\begin{eqnarray}
\label{E2} \vert\Psi(t)\rangle&=&\alpha\vert\textbf{0}\rangle+
\beta[\cos(\Theta_r t)\vert\textbf{1}_1\rangle-i\sin(\Theta_r
t)\vert\textbf{1}_N\rangle].
\end{eqnarray}
At the moment $\Theta_r t_f=\pi/2$ they turn off the couplings and
Bob gets the state
$\vert\Psi(t_f)\rangle=\alpha\vert\textbf{0}\rangle-i\beta\vert\textbf{1}_N\rangle$.
If Bob performs a gate operation  $U=(1,i)$, he could retrieve the
state $\alpha\vert0\rangle_N+\beta\vert1\rangle_N$ for the atom $N$.
The procedure completes QIT inside the one-dimensional coupled
resonator waveguide, which in principle could be extended to realize
short distance quantum communications. Different from the schemes
based on spin chains for short distance quantum communications, the
principle advantage of this scheme is that, in the coupled-resonator
waveguide, individual lattices sites can be addressed with optical
lasers. Therefore, it is much easier to switch of interactions of
the qubit on which the initial state is encoded and the qubit on
which the final state is received with the rest of the waveguide in
this proposal.

It is noted that QIT can also be implemented through STIRAP
techniques \cite{RMP-70-1003} with this one-dimensional coupled
resonator waveguide. In such a case, we require the lasers to select
a resonant transition from the initial state
$\vert\textbf{1}_1\rangle$ to the final state
$\vert\textbf{1}_N\rangle$ via an intermediate state such as
$\vert\tilde{\textbf{k}}\rangle$, while other transition channels
are far off resonance. Then the system is reduced to a typical
$\Lambda$ configuration, which supports a dark state involving the
two states $\vert\textbf{1}_1\rangle$ and
$\vert\textbf{1}_N\rangle$. Adiabatic passage following the dark
state can be implemented by varying the Rabi frequencies slowly.
Then an arbitrary unknown state $\alpha\vert 0\rangle_1+\beta\vert
1\rangle_1$ can be transferred directly from the first atom to the
end one following the STIRAP.

It is necessary to verify the model through numerical simulations.
We consider the case of QIT in three coupled resonators. The system
is initially prepared in the state
$\frac{1}{\sqrt{2}}(\vert0\rangle_1+\vert1\rangle_1)\vert0\rangle_2\vert0\rangle_3$.
Employing a quantum master equation approach, we have simulated the
dynamics of the system through the Monte Carlo wave function (MCWF)
formalism \cite{RMP-70-101,cpc}. In Fig. 3 the numerical solutions
of the density matrix equations for the full system described by the
exact Hamiltonian $H$ are shown together with the dynamics of the
system undergoing the effective Hamiltonian (\ref{H3}). Here the
parameters are chosen such that they are within the parameter range
for which the scheme is valid (discussed in the next paragraph). It
is clear that the agreement between the exact and effective model is
excellent under the given parameters. The system starts from the
state
$\frac{1}{\sqrt{2}}(\vert0\rangle_1+\vert1\rangle_1)\vert0\rangle_2\vert0\rangle_3$.
At the time $t=\pi/2\Theta_r$, the first atom evolves into its
ground state $\vert0_1\rangle$ and the third atom evolves into
$\frac{1}{\sqrt{2}}(\vert0\rangle_3-i\vert1\rangle_3)$. This process
completes the procedure for QIT between these two atoms. During this
process, the populations of the atomic excited states and the cavity
modes keep small.

We now study the performance of this protocol under realistic
circumstances and estimate the range of parameters implementing
optimal QIT.
\begin{figure}[h]
\centering
\includegraphics[bb=4 4 220 216,totalheight=3in,clip]{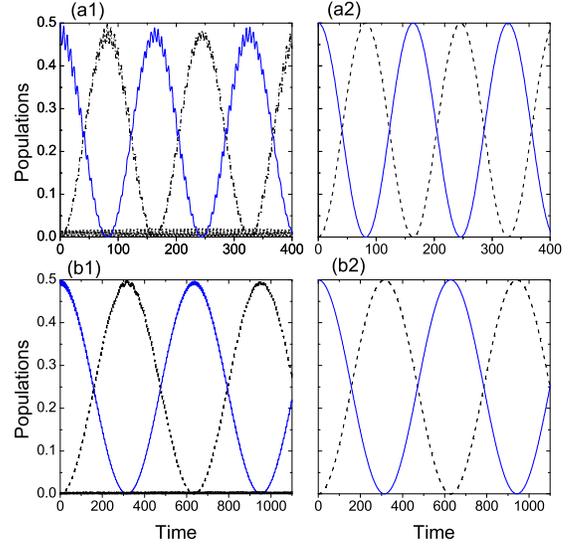}
\caption{\label{F3}Evolution of the system from both exact
calculations of the master equation (a1, b1) and the solutions for
the effective Hamiltonian (\ref{H3}) (a2, b2). In all the Figures,
Solid lines represent the population of
$\vert1\rangle_1\vert0\rangle_2\vert0\rangle_3$, Dot lines the
population of $\vert0\rangle_1\vert0\rangle_2\vert1\rangle_3$. The
parameters are chosen as $\Delta=10g,J_c=0.5g$, for (a1), (a2),
$\Omega_1=\Omega_3=0.02g$, and for (b1), (b2),
$\Omega_1=\Omega_3=0.01g$. Time is measured in unit of $g^{-1}$.}
\end{figure}
Consider a one-dimensional coupled resonator waveguide consisting of
$N$ nodes. Alice wants to transfer an arbitrary quantum state from
the first node to Bob who is located at the end. To quantify the
performance of QIT, we utilize the fidelity $F=\langle \psi_p \vert
\rho_f\vert \psi_p\rangle$, where $\vert \psi_p\rangle$ refers to
the perfectly transferred state, and $\rho_f$ is the final reduced
density matrix of the last atom under realistic circumstances. The
fidelity is reduced due to the small probabilities of populating
either the atomic excited states or the cavity modes. For this
protocol, spontaneous emission from the state $\vert e\rangle_j$ at
a rate $\gamma$ and cavity decay of photons at a rate $\kappa$ lead
to effective decay rates $\Gamma_E=\sum_k|\Omega_k/E_k|^2\gamma$ and
$\Gamma_C=\sum_k|g/(\sqrt{N}\delta_k)|^2\kappa$, with
$\Omega_k=\mbox{max}(\Omega_{1k},\Omega_{2k})$. Hence to achieve
coherent interaction requires that $\{\Gamma_E,\Gamma_C\}<
\Theta_r$. These requirements could be satisfied if $\gamma\ll
J_Cg^2/\Delta^2$ and $\kappa\ll J_C$. Since photons are more likely
to tunnel to the next cavity than decay into free space, $\kappa\ll
J_C$ should hold in most cases. For the condition $\gamma\ll
J_Cg^2/\Delta^2$ to hold, cavities with a high ratio $g/\gamma$ are
very good candidates. These two requirements together imply that the
cavities should have a high cooperativity factor. To make sure this
scheme is valid, we also require that $\Delta\gg g$ and
$g^2/\Delta\gg\Omega_i$. Taking into account these probabilities of
error, the fidelity is estimated as
$F\simeq1-\Gamma_Et_f-\Gamma_Ct_f$, where $t_f=\pi/2\Theta_r$ is the
time to complete QIT.

For experimental implementation of QIT in a coupled resonator
waveguide, atoms or polar molecules trapped in coupled
superconducting stripline microwave resonators
\cite{nature-451-664,Prl-100-170501} are promising candidates. It is
noted that hybrid devices combining solid state circuits with
trapped atoms or molecules have been explored \cite{Prl-97-033003}.
\begin{figure}[h]
\centering
\includegraphics[bb=34 58 522 434,totalheight=1.7in,clip]{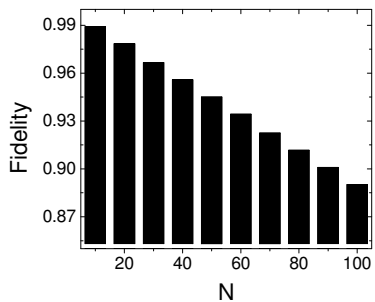}
\caption{\label{F4}Fidelity bars for different waveguide lengths
$N$. Parameters are given in the text.}
\end{figure}
We choose the parameters as $g\sim2\pi\times200$ MHz,
$\Delta\sim2\pi\times2$ GHz, $J_C\sim2\pi\times100$ MHz,
$\gamma\sim2\pi\times20$ kHz, $\kappa\sim2\pi\times50$ kHz
\cite{Prl-100-170501}, and $\Omega\sim2\pi\times2$ MHz. Then we can
estimate the fidelity of this state transfer channel. In Fig. 4 we
display the fidelity for various waveguide lengths $N$. We see that
as the cavity number increases the fidelity decreases. For a
waveguide consisted of 100 coupled resonators, the fidelity is about
$88\%$ and the time to complete QIT is $t_f\sim0.01$ $\mu s$. Thus
the number of cavities should be within 100 to make this scheme
efficient. To improve the fidelity and correct the error for QIT in
this network, the proposed methods for quantum error correction can
be utilized \cite{RMP-74-347}.

\section{CONCLUSION}  We have presented a protocol for the
implementation of short distance quantum communications in a
one-dimensional coupled-resonator waveguide. This protocol utilizes
the cavity field induced nonlocal interactions and Raman transitions
between trapped atoms at the opposite ends of the waveguide. QIT
could take place directly from the first node to the end one without
involving the intermediate nodes, which represents an interesting
step towards realizing quantum communications.

\section*{ACKNOWLEDGMENTS}
This work was supported by the National Natural Science Foundation
of China under Grants Nos. 10674009, 10874004, 10821062 and National
Key Basic Research Program No.2006CB921601. Peng-Bo Li acknowledges
the quite useful discussions with Hong-Yan Li.

%\newpage
%\bibliography{manuscript}

\begin{thebibliography}{27}
\expandafter\ifx\csname
natexlab\endcsname\relax\def\natexlab#1{#1}\fi
\expandafter\ifx\csname bibnamefont\endcsname\relax
  \def\bibnamefont#1{#1}\fi
\expandafter\ifx\csname bibfnamefont\endcsname\relax
  \def\bibfnamefont#1{#1}\fi
\expandafter\ifx\csname citenamefont\endcsname\relax
  \def\citenamefont#1{#1}\fi
\expandafter\ifx\csname url\endcsname\relax
  \def\url#1{\texttt{#1}}\fi
\expandafter\ifx\csname urlprefix\endcsname\relax\def\urlprefix{URL
}\fi \providecommand{\bibinfo}[2]{#2}
\providecommand{\eprint}[2][]{\url{#2}}

\bibitem[{\citenamefont{Nielsen and Chuang}(2000)}]{quantum_information}
\bibinfo{author}{\bibfnamefont{M.~A.} \bibnamefont{Nielsen}} \bibnamefont{and}
  \bibinfo{author}{\bibfnamefont{I.~L.} \bibnamefont{Chuang}},
  \emph{\bibinfo{title}{Quantum Computation and Quantum Information}}
  (\bibinfo{publisher}{Cambridge University Press, Cambridge, UK},
  \bibinfo{year}{2000}).

\bibitem[{\citenamefont{Kimble}(2008)}]{nature-453-1023}
\bibinfo{author}{\bibfnamefont{H.~J.} \bibnamefont{Kimble}},
  \bibinfo{journal}{Nature (London)} \textbf{\bibinfo{volume}{453}},
  \bibinfo{pages}{1023} (\bibinfo{year}{2008}).

\bibitem[{\citenamefont{Kimble}(1998)}]{Kimble}
\bibinfo{author}{\bibfnamefont{H.~J.} \bibnamefont{Kimble}},
  \bibinfo{journal}{Phys.\ Scr.} \textbf{\bibinfo{volume}{T76}},
  \bibinfo{pages}{127} (\bibinfo{year}{1998}).

\bibitem[{\citenamefont{Mabuchi and Doherty}(2002)}]{Sci298}
\bibinfo{author}{\bibfnamefont{H.}~\bibnamefont{Mabuchi}} \bibnamefont{and}
  \bibinfo{author}{\bibfnamefont{A.~C.} \bibnamefont{Doherty}},
  \bibinfo{journal}{Science} \textbf{\bibinfo{volume}{298}},
  \bibinfo{pages}{1372} (\bibinfo{year}{2002}).

\bibitem[{\citenamefont{Cirac et~al.}(1997)\citenamefont{Cirac, Zoller, Kimble,
  and Mabuchi}}]{prl-78-3221}
\bibinfo{author}{\bibfnamefont{J.~I.} \bibnamefont{Cirac}},
  \bibinfo{author}{\bibfnamefont{P.}~\bibnamefont{Zoller}},
  \bibinfo{author}{\bibfnamefont{H.~J.} \bibnamefont{Kimble}},
  \bibnamefont{and} \bibinfo{author}{\bibfnamefont{H.}~\bibnamefont{Mabuchi}},
  \bibinfo{journal}{Phys.\ Rev. Lett.} \textbf{\bibinfo{volume}{78}},
  \bibinfo{pages}{3221} (\bibinfo{year}{1997}).

\bibitem[{\citenamefont{Boozer et~al.}(2007)\citenamefont{Boozer, Boca, Miller,
  Northup, and Kimble}}]{prl-98-193601}
\bibinfo{author}{\bibfnamefont{A.~D.} \bibnamefont{Boozer}},
  \bibinfo{author}{\bibfnamefont{A.}~\bibnamefont{Boca}},
  \bibinfo{author}{\bibfnamefont{R.}~\bibnamefont{Miller}},
  \bibinfo{author}{\bibfnamefont{T.~E.} \bibnamefont{Northup}},
  \bibnamefont{and} \bibinfo{author}{\bibfnamefont{H.~J.}
  \bibnamefont{Kimble}}, \bibinfo{journal}{Phys.\ Rev. Lett.}
  \textbf{\bibinfo{volume}{98}}, \bibinfo{pages}{193601}
  (\bibinfo{year}{2007}).

\bibitem[{\citenamefont{Bose}(2003)}]{prl-91-207901}
\bibinfo{author}{\bibfnamefont{S.}~\bibnamefont{Bose}},
  \bibinfo{journal}{Phys.\ Rev. Lett.} \textbf{\bibinfo{volume}{91}},
  \bibinfo{pages}{207901} (\bibinfo{year}{2003}).

\bibitem[{\citenamefont{Hartmann et~al.}(2006)\citenamefont{Hartmann, Brandao,
  and Plenio}}]{Nature-physics-2-849}
\bibinfo{author}{\bibfnamefont{M.~J.} \bibnamefont{Hartmann}},
  \bibinfo{author}{\bibfnamefont{F.~G. S.~L.} \bibnamefont{Brandao}},
  \bibnamefont{and} \bibinfo{author}{\bibfnamefont{M.~B.}
  \bibnamefont{Plenio}}, \bibinfo{journal}{Nature Phys.}
  \textbf{\bibinfo{volume}{2}}, \bibinfo{pages}{849} (\bibinfo{year}{2006}).

\bibitem[{\citenamefont{Greentree et~al.}(2006)\citenamefont{Greentree, Tanhan,
  Cole, and Hollenberg}}]{Nature-physics-2-856}
\bibinfo{author}{\bibfnamefont{A.~D.} \bibnamefont{Greentree}},
  \bibinfo{author}{\bibfnamefont{C.}~\bibnamefont{Tanhan}},
  \bibinfo{author}{\bibfnamefont{J.~H.} \bibnamefont{Cole}}, \bibnamefont{and}
  \bibinfo{author}{\bibfnamefont{L.~C.~L.} \bibnamefont{Hollenberg}},
  \bibinfo{journal}{Nature Phys.} \textbf{\bibinfo{volume}{2}},
  \bibinfo{pages}{856} (\bibinfo{year}{2006}).

\bibitem[{\citenamefont{Hartmann et~al.}(2007)\citenamefont{Hartmann, Brandao,
  and Plenio}}]{Prl-99-160501}
\bibinfo{author}{\bibfnamefont{M.~J.} \bibnamefont{Hartmann}},
  \bibinfo{author}{\bibfnamefont{F.~G. S.~L.} \bibnamefont{Brandao}},
  \bibnamefont{and} \bibinfo{author}{\bibfnamefont{M.~B.}
  \bibnamefont{Plenio}}, \bibinfo{journal}{Phys.\ Rev. Lett.}
  \textbf{\bibinfo{volume}{99}}, \bibinfo{pages}{160501}
  (\bibinfo{year}{2007}).

\bibitem[{\citenamefont{Hartmann and Plenio}(2007)}]{Prl-99-103601}
\bibinfo{author}{\bibfnamefont{M.~J.} \bibnamefont{Hartmann}} \bibnamefont{and}
  \bibinfo{author}{\bibfnamefont{M.~B.} \bibnamefont{Plenio}},
  \bibinfo{journal}{Phys.\ Rev. Lett.} \textbf{\bibinfo{volume}{99}},
  \bibinfo{pages}{103601} (\bibinfo{year}{2007}).

\bibitem[{\citenamefont{Ji et~al.}(2007)\citenamefont{Ji, Xie, and
  Liu}}]{Prl-99-183602}
\bibinfo{author}{\bibfnamefont{A.-C.} \bibnamefont{Ji}},
  \bibinfo{author}{\bibfnamefont{X.~C.} \bibnamefont{Xie}}, \bibnamefont{and}
  \bibinfo{author}{\bibfnamefont{W.~M.} \bibnamefont{Liu}},
  \bibinfo{journal}{Phys.\ Rev. Lett.} \textbf{\bibinfo{volume}{99}},
  \bibinfo{pages}{183602} (\bibinfo{year}{2007}).

\bibitem[{\citenamefont{Zhou et~al.}(2008)\citenamefont{Zhou, Gong, Liu, Sun,
  and Nori}}]{Prl-101-100501}
\bibinfo{author}{\bibfnamefont{L.}~\bibnamefont{Zhou}},
  \bibinfo{author}{\bibfnamefont{Z.~R.} \bibnamefont{Gong}},
  \bibinfo{author}{\bibfnamefont{Y.~X.} \bibnamefont{Liu}},
  \bibinfo{author}{\bibfnamefont{C.~P.} \bibnamefont{Sun}}, \bibnamefont{and}
  \bibinfo{author}{\bibfnamefont{F.}~\bibnamefont{Nori}},
  \bibinfo{journal}{Phys.\ Rev. Lett.} \textbf{\bibinfo{volume}{101}},
  \bibinfo{pages}{100501} (\bibinfo{year}{2008}).

\bibitem[{\citenamefont{Cho et~al.}(2008)\citenamefont{Cho, Angelakis, and
  Bose}}]{Prl-101-246809}
\bibinfo{author}{\bibfnamefont{J.}~\bibnamefont{Cho}},
  \bibinfo{author}{\bibfnamefont{D.~G.} \bibnamefont{Angelakis}},
  \bibnamefont{and} \bibinfo{author}{\bibfnamefont{S.}~\bibnamefont{Bose}},
  \bibinfo{journal}{Phys.\ Rev. Lett.} \textbf{\bibinfo{volume}{101}},
  \bibinfo{pages}{246809} (\bibinfo{year}{2008}).

\bibitem[{\citenamefont{Li et~al.}()\citenamefont{Li, Gu, Gong, and Guo}}]{li}
\bibinfo{author}{\bibfnamefont{P.~B.} \bibnamefont{Li}},
  \bibinfo{author}{\bibfnamefont{Y.}~\bibnamefont{Gu}},
  \bibinfo{author}{\bibfnamefont{Q.~H.} \bibnamefont{Gong}}, \bibnamefont{and}
  \bibinfo{author}{\bibfnamefont{G.~C.} \bibnamefont{Guo}}, \eprint{Eur. Phys.
  J. D, submitted}.

\bibitem[{\citenamefont{Bergmann et~al.}(1998)\citenamefont{Bergmann, Theuer,
  and Shore}}]{RMP-70-1003}
\bibinfo{author}{\bibfnamefont{K.}~\bibnamefont{Bergmann}},
  \bibinfo{author}{\bibfnamefont{H.}~\bibnamefont{Theuer}}, \bibnamefont{and}
  \bibinfo{author}{\bibfnamefont{B.~W.} \bibnamefont{Shore}},
  \bibinfo{journal}{Rev.\ Mod.\ Phys.} \textbf{\bibinfo{volume}{70}},
  \bibinfo{pages}{1003} (\bibinfo{year}{1998}).

\bibitem[{\citenamefont{Bennett et~al.}(1993)\citenamefont{Bennett, Brassard,
  Crepeau, Jozsa, Peres, and Wootters}}]{Prl-70-1895}
\bibinfo{author}{\bibfnamefont{C.~H.} \bibnamefont{Bennett}},
  \bibinfo{author}{\bibfnamefont{G.}~\bibnamefont{Brassard}},
  \bibinfo{author}{\bibfnamefont{C.}~\bibnamefont{Crepeau}},
  \bibinfo{author}{\bibfnamefont{R.}~\bibnamefont{Jozsa}},
  \bibinfo{author}{\bibfnamefont{A.}~\bibnamefont{Peres}}, \bibnamefont{and}
  \bibinfo{author}{\bibfnamefont{W.~K.} \bibnamefont{Wootters}},
  \bibinfo{journal}{Phys.\ Rev. Lett.} \textbf{\bibinfo{volume}{70}},
  \bibinfo{pages}{1895} (\bibinfo{year}{1993}).

\bibitem[{\citenamefont{Cleve and Buhrman}(1997)}]{pra-56-1201}
\bibinfo{author}{\bibfnamefont{R.}~\bibnamefont{Cleve}} \bibnamefont{and}
  \bibinfo{author}{\bibfnamefont{H.}~\bibnamefont{Buhrman}},
  \bibinfo{journal}{Phys.\ Rev. A} \textbf{\bibinfo{volume}{56}},
  \bibinfo{pages}{1201} (\bibinfo{year}{1997}).

\bibitem[{\citenamefont{Hennessy et~al.}(2007)\citenamefont{Hennessy, Badolato,
  Winger, Gerace, Atatue, Gulde, Falt, Hu, and Imamoglu}}]{nature-445-896}
\bibinfo{author}{\bibfnamefont{K.}~\bibnamefont{Hennessy}},
  \bibinfo{author}{\bibfnamefont{A.}~\bibnamefont{Badolato}},
  \bibinfo{author}{\bibfnamefont{M.}~\bibnamefont{Winger}},
  \bibinfo{author}{\bibfnamefont{D.}~\bibnamefont{Gerace}},
  \bibinfo{author}{\bibfnamefont{M.}~\bibnamefont{Atatue}},
  \bibinfo{author}{\bibfnamefont{S.}~\bibnamefont{Gulde}},
  \bibinfo{author}{\bibfnamefont{S.}~\bibnamefont{Falt}},
  \bibinfo{author}{\bibfnamefont{E.~L.} \bibnamefont{Hu}}, \bibnamefont{and}
  \bibinfo{author}{\bibfnamefont{A.}~\bibnamefont{Imamoglu}},
  \bibinfo{journal}{Nature (London)} \textbf{\bibinfo{volume}{445}},
  \bibinfo{pages}{896} (\bibinfo{year}{2007}).

\bibitem[{\citenamefont{Schoelkopf and Girvin}(2008)}]{nature-451-664}
\bibinfo{author}{\bibfnamefont{R.~J.} \bibnamefont{Schoelkopf}}
  \bibnamefont{and} \bibinfo{author}{\bibfnamefont{S.~M.}
  \bibnamefont{Girvin}}, \bibinfo{journal}{Nature (London)}
  \textbf{\bibinfo{volume}{451}}, \bibinfo{pages}{664} (\bibinfo{year}{2008}).

\bibitem[{\citenamefont{Zheng and Guo}(2000)}]{Guo}
\bibinfo{author}{\bibfnamefont{S.-B.} \bibnamefont{Zheng}} \bibnamefont{and}
  \bibinfo{author}{\bibfnamefont{G.-C.} \bibnamefont{Guo}},
  \bibinfo{journal}{Phys.\ Rev. Lett.} \textbf{\bibinfo{volume}{85}},
  \bibinfo{pages}{2392} (\bibinfo{year}{2000}).

\bibitem[{\citenamefont{James}(2000)}]{method}
\bibinfo{author}{\bibfnamefont{D.~F.~V.} \bibnamefont{James}},
  \bibinfo{journal}{Fortschr. Phys.} \textbf{\bibinfo{volume}{48}},
  \bibinfo{pages}{823} (\bibinfo{year}{2000}).

\bibitem[{\citenamefont{Plenio and Knight}(1998)}]{RMP-70-101}
\bibinfo{author}{\bibfnamefont{M.~B.} \bibnamefont{Plenio}} \bibnamefont{and}
  \bibinfo{author}{\bibfnamefont{P.~L.} \bibnamefont{Knight}},
  \bibinfo{journal}{Rev.\ Mod.\ Phys.} \textbf{\bibinfo{volume}{70}},
  \bibinfo{pages}{101} (\bibinfo{year}{1998}).

\bibitem[{\citenamefont{Schack and Brun}(1997)}]{cpc}
\bibinfo{author}{\bibfnamefont{R.}~\bibnamefont{Schack}} \bibnamefont{and}
  \bibinfo{author}{\bibfnamefont{T.~A.} \bibnamefont{Brun}},
  \bibinfo{journal}{Comput.\ Phys.\ Commun.} \textbf{\bibinfo{volume}{102}},
  \bibinfo{pages}{210} (\bibinfo{year}{1997}).

\bibitem[{\citenamefont{Petrosyan and Fleischhauer}(2008)}]{Prl-100-170501}
\bibinfo{author}{\bibfnamefont{D.}~\bibnamefont{Petrosyan}} \bibnamefont{and}
  \bibinfo{author}{\bibfnamefont{M.}~\bibnamefont{Fleischhauer}},
  \bibinfo{journal}{Phys.\ Rev. Lett.} \textbf{\bibinfo{volume}{100}},
  \bibinfo{pages}{170501} (\bibinfo{year}{2008}).

\bibitem[{\citenamefont{Rabl et~al.}(2006)\citenamefont{Rabl, DeMille, Doyle,
  Lukin, Schoelkopf, and Zoller}}]{Prl-97-033003}
\bibinfo{author}{\bibfnamefont{P.}~\bibnamefont{Rabl}},
  \bibinfo{author}{\bibfnamefont{D.}~\bibnamefont{DeMille}},
  \bibinfo{author}{\bibfnamefont{J.~M.} \bibnamefont{Doyle}},
  \bibinfo{author}{\bibfnamefont{M.~D.} \bibnamefont{Lukin}},
  \bibinfo{author}{\bibfnamefont{R.~J.} \bibnamefont{Schoelkopf}},
  \bibnamefont{and} \bibinfo{author}{\bibfnamefont{P.}~\bibnamefont{Zoller}},
  \bibinfo{journal}{Phys.\ Rev. Lett.} \textbf{\bibinfo{volume}{97}},
  \bibinfo{pages}{033003} (\bibinfo{year}{2006}).

\bibitem[{\citenamefont{Galindo and Martin-Delgado}(2002)}]{RMP-74-347}
\bibinfo{author}{\bibfnamefont{A.}~\bibnamefont{Galindo}} \bibnamefont{and}
  \bibinfo{author}{\bibfnamefont{M.~A.} \bibnamefont{Martin-Delgado}},
  \bibinfo{journal}{Rev.\ Mod.\ Phys.} \textbf{\bibinfo{volume}{74}},
  \bibinfo{pages}{347} (\bibinfo{year}{2002}).

\end{thebibliography}

\end{document}